\documentstyle[eqsecnum,preprint,aps]{revtex}
\def\beq{\begin{equation}}
\def\eeq{\end{equation}}
\begin{document}

\draft

\title{Extended RPA within a solvable 3 level model}

\author{F. Catara $^{1,2}$$\thanks{e-mail address: catara@ct.infn.it}$,
M. Grasso $^{3}$$\thanks{e-mail address: grasso@ipno.in2p3.fr}$, 
M. Sambataro $^{2}$$\thanks{e-mail address: samba@ct.infn.it}$}
\address{$^{1}$ Dipartimento di Fisica, Universit\`a di Catania}
\address{$^{2}$ Istituto Nazionale di Fisica Nucleare, Sez. di Catania}
\address{Corso Italia 57, I-95129 Catania, Italy}
\address{$^{3}$ Institut de Physique Nucl\'eaire, IN2P3-CNRS, \\
Universit\'e
Paris-Sud, 91406 Orsay Cedex, France}
\maketitle

\begin{abstract}
Working within an exactly solvable 3 level model,
we discuss an extension of the Random Phase Approximation (RPA) based
on a boson formalism. A boson Hamiltonian is defined via a mapping
procedure and its expansion truncated at four-boson terms.
RPA-type equations are then constructed and solved iteratively.
The new solutions gain in stability with respect to the RPA ones.
We perform diagonalizations of the boson Hamiltonian in spaces 
containing up to four-phonon components. Approximate spectra exhibit
an improved quality with increasing the size of these multiphonon
spaces. Special attention is addressed to the problem of the
anharmonicity of the spectrum.
\end{abstract}


\section{Introduction}
The most commonly used microscopic approach for the study of collective
vibrational states in many-fermion systems is the Random Phase Approximation
(RPA)\cite{RS}. In this theory the lowest collective excitations result
from the action of phonon operators $Q^{\dagger}_{\nu}$ on a state $|RPA)$
which is defined by the condition that $Q_{\nu}|RPA)=0$. This state represents
the ground state of the system. It is a distinctive feature of RPA
that multiphonon states, i.e. states obtained by repeated actions
of phonon operators on the ground state, are eigenstates of the Hamiltonian 
with energies forming a harmonic spectrum. The existence of
states which can be approximately described as corresponding to the
multiple excitation of low-lying and/or high-lying phonons is well established 
in atomic nuclei. However, deviations from the harmonic picture are also 
observed and their influence on several processes has been 
analysed\cite{harakeh}.

In a standard derivation of the RPA equations, a crucial point is represented
by the so-called Quasi-Boson Approximation (QBA). This is a rather crude
approximation which causes the $Q^{\dagger}_{\nu}$ operators to behave as
boson operators in spite of their (composite) fermionic structure.
Overcoming this approximation has represented the starting point of many
attempts aiming at improving RPA\cite{duke,schuck,klein3,kara,cata,duke2,%
cata2,radu,grasso,krmp,simko1,gra,dang,chomaz,sambac,samba2,ccg3,%
lanza,volpe,dinh}. One of the line of research in such
a context has been based on a reformulation of the whole theory in a boson
formalism\cite{chomaz,sambac,samba2,ccg3,lanza,volpe,dinh}.
In other words, the $Q^{\dagger}_{\nu}$ operators have been
defined  from the beginning in terms of true boson operators and all the
fermion operators of interest have been replaced, via a mapping procedure,
by their boson images. The RPA-type equations that one constructs in this
formalism depend, of course, on the degree of expansion of the boson
Hamiltonian. Truncating this expansion at the lowest order, i.e. at
two-boson terms only, gives the boson counterpart of RPA. Including 
higher-order terms in the boson image of the Hamiltonian provides a natural
way to reach a higher level of approximation. Besides that, the inclusion
of these terms has another important effect: it leads to a coupling
among multiphonon states. States which result from a diagonalization in
a m-phonon space are therefore superpositions of zero-, one-,... , m-phonon
states. Such a diagonalization is expected to lead to a
further improved degree of approximation as well as to cause anharmonicities
in the spectrum.

Calculations in this boson formalism have been performed in the recent
past for atomic nuclei\cite{ccg3,lanza}
considering a Hamiltonian truncated at four-boson terms and diagonalizing it
in the space of one- and two-phonon states.
The resulting anharmonicities have not been found large, especially in 
$^{208}$Pb. 
In particular, the anharmonicity associated 
with states whose main component is a Double Giant Resonance has been 
found of the order
of a few hundred $KeV$. This is certainly related to the fact that RPA gives
a good description of Giant Resonances, especially in heavy closed shell
nuclei. 

In metallic clusters, a strongly collective state is known to exist, the dipole
plasmon, which corresponds to the oscillation of the delocalized electrons of
the cluster against the positively charged ions. The experimental evidence
for states corresponding to the double excitation of the plasmon
has not been confirmed \cite{haba}. From the theoretical point of view the
situation is also quite unclear. From one hand, in \cite{h} a purely
harmonic spectrum for the multiple excitation of the plasmon has been predicted.
On the other hand, by using the same approach as in \cite{ccg3,lanza} huge 
anharmonicities in the two-plasmon states have been found \cite{ccg2}. An 
important difference with respect to the case of atomic nuclei is that in
metallic clusters the two-body interaction is very long ranged. This is 
probably the main reason why the RPA ground state is very different from the 
Hartree-Fock one and the RPA backward amplitudes are quite large. 
Of course, this may cause that the same level of truncation in the boson
expansion is not adequate both in the case of nuclei and of metal 
clusters.

In principle, configuration mixing calculations can give a clear, model
independent, indication on the existence of such two-phonon states and on
their degree of anharmonicity. Unfortunately, since the states one looks
for are quite high in energy, the number of configurations required to get
stable results is huge. In \cite{two} such a study has been performed for a
very simple case: two interacting electrons moving in a uniform positive
charge distribution. This is a kind of precursor of a metal cluster in
the jellium approximation and allows for a numerically exact calculation.
Important deviations from the harmonic limit have been found. More specifically,
in addition to an almost perfectly harmonic vibrational band based on the
ground state, the other states can be classified in other bands with a much
less degree of harmonicity.

Aim of this paper is to shed some light on the limits of the approach adopted
in \cite{lanza} and \cite{ccg2} by 
applying it to a 3-level solvable model\cite{lip3}.
The analysis we are
going to present is very similar to that in \cite{volpe} where a 2-level
model was considered and the parameters were adjusted in such a way to
mimick the multiple excitation of a Giant Resonance. The 3-level model is,
of course, richer. In particular, since there are two single particle states
above the Fermi surface (particle states) and one below (hole),  
two different elementary p-h configurations and, correspondingly, two different 
phonons can be excited. Therefore, one can better simulate the situation 
encountered in nuclei 
which generally present one high-lying and one low-lying collective modes for 
each multipolarity. Also, matrix elements of the interaction connecting a 
particle-hole state  with a two-particle one can be included in a natural way.
These terms are present in a generic two-body interaction and are very 
important since they couple states having numbers of phonons differing by one.

The paper is organized as follows. In Sec. II, we will describe the model
and the formalism. In Sec. III, we will analyse the results. Finally,
in Sec. IV, we will review the content of the paper and draw some conclusions.

\section{The model and the formalism}
The model\cite{lip3} consists of three $2\Omega$-fold degenerate
single-particle shells which are occupied by $2\Omega$
particles. In the absence of interaction, then, the lowest level is
completely filled while the others are empty. This state,
the ``Hartree-Fock" (HF) state of the system, is denoted by $|0\rangle$.
A single-particle state is specified by a set
of quantum numbers ($j,m$), where $j$ stands for the shell
($j$=0,1,2) and $m$ specifies the $2\Omega$ substates within the shell.
The creation and annihilation operators of a fermion in a state ($j,m$)
are defined by $a^{\dagger}_{jm}$ and $a_{jm}$, respectively.

Let us consider the operators
\begin{equation}
K_{ij}=\sum^{2\Omega}_{m=1}a^{\dagger}_{im}a_{jm}~~~~(i,j=0,1,2).
\label{1}
\end{equation}
These operators satisfy the Lie algebra of the group SU(3)
\begin{equation}
[K_{ij},K_{kl}]=\delta_{jk}K_{il}-\delta_{il}K_{kj}.
\label{2}
\end{equation}
The Hamiltonian of the model 
is written in terms of the generators
$K_{ij}$ only and contains up to two-body interactions. Its form is
\begin{eqnarray}
H_F=&&\sum_{i=1,2}\epsilon (i)K_{ii}+%
      \sum_{i,j=1,2}V_x(i,j)K_{i0}K_{0j}\nonumber\\
    &&+\frac{1}{2}\sum_{i,j=1,2}V_v(i,j)(K_{i0}K_{j0}+K_{0j}K_{0i})\nonumber\\
    &&+\sum_{i,j,k=1,2}V_y(i,j,k)(K_{i0}K_{jk}+K_{kj}K_{0i}),
\label{3}
\end{eqnarray}
with real coefficients. The eigenstates of $H_F$ can be
constructed either by using the properties of the SU(3) algebra or 
by diagonalizing it in the space
\begin{equation}
F=\left\{|n_1n_2\rangle =%
\frac{1}{\sqrt{{\cal N}_{n_1n_2}}}(K_{10})^{n_1}(K_{20})^{n_2}%
|0\rangle\right\}_{0\leq n_1+n_2\leq 2\Omega},
\label{4}
\end{equation}
where ${\cal N}_{n_1n_2}$ are normalization factors.

In order to introduce the boson formalism, let's 
define the space
\begin{equation}
B=\left\{|n_1n_2) =\frac{1}{\sqrt{n_1!n_2!}}%
(b^{\dagger}_1)^{n_1}(b^{\dagger}_2)^{n_2}%
|0)\right\}_{0\leq n_1+n_2\leq 2\Omega},
\label{5}
\end{equation}
where the operators
$b^{\dagger}_i$ obey the standard boson commutation relations
\begin{equation}
[b_i,b^{\dagger}_j]=\delta _{ij},~~~~~[b_i,b_j]=0
\label{6}
\end{equation}
and $|0)$ is the vacuum of the $b_i$'s operators.
A one-to-one correspondence exists between the states of $F$ and $B$, the
boson operators $b^{\dagger}_i$ playing the role of the excitation operators
$K_{i0}$ and the boson vacuum $|0)$ replacing the HF state
$|0\rangle$. 

The mapping procedure to construct boson images of fermion operators is
the same discussed in 
previous works (see, for instance, Ref. \cite{samba3}) and, due to the 
orthonormality of both sets of states
$|n_1,n_2\rangle$ and $|n_1,n_2)$, it is simply based on the requirement that
corresponding matrix elements in $F$ and $B$ be equal. The procedure
is, therefore, of Marumori-type. We refer to Ref. \cite{samba3} for more 
details. Here, we simply say that, in correspondence
with the Hamiltonian $H_F$ (\ref{3}), we introduce 
a hermitian boson Hamiltonian $H_B$ which, in the most extended version,
has the form
\begin{eqnarray}
H_B=&&\alpha +\sum_i\beta_i(b^{\dagger}_i+h.c.)+%
      \sum_{ij}\gamma_{ij}b^{\dagger}_ib_j+%
      \sum_{i\leq j}\phi_{ij}(b^{\dagger}_ib^{\dagger}_j+h.c.)\nonumber\\
    &&+\sum_{i\leq j}\sum_k\epsilon_{ijk}(b^{\dagger}_ib^{\dagger}_j%
      b_k+h.c.)+%
      \sum_{i\leq j}\sum_{k\leq l}\delta_{ijkl}%
      b^{\dagger}_ib^{\dagger}_jb_kb_l.
\label{7}
\end{eqnarray}
with the coefficients depending on the parameters 
$\epsilon (i), V_x, V_v, V_y$ of Eq.(\ref{3}).

Let us now introduce the operators 
\begin{equation}
Q^{\dagger}_{\nu}=\sum_iX^{(\nu )}_ib^{\dagger}_i-%
\sum_iY^{(\nu )}_ib_i.
\label{8}
\end{equation}
and let the state $|\Psi_0)$ satisfy the condition
\begin{equation}
Q_{\nu}|\Psi_0)=0.
\label{9}
\end{equation}
By using the equations of motion method \cite{rowe} one finds that the
amplitudes X and Y are solutions of
\begin{equation}	
\left(\begin{array}{c}
~A~~~~~~~B\\-B^*~~-A^*\end{array}\right)
\left(\begin{array}{c}
X^{(\nu)}\\Y^{(\nu)}\end{array}\right)=\omega^{(\nu)}
\left(\begin{array}{c}
X^{(\nu)}\\Y^{(\nu)}\end{array}\right)\
\label{10}
\end{equation}
where 
\begin{equation}
A_{ij}=(\Psi_0|[b_i,[H_B,b^{\dagger}_j]]|\Psi_0) ,
\label{11}
\end{equation}
\begin{equation}
B_{ij}=-(\Psi_0|[b_i,[H_B,b_j]]|\Psi_0)
\label{12}
\end{equation}
and $\omega^{(\nu)}$ are the energies of the excited states 
\begin{equation}
|\nu )=Q^{\dagger}_{\nu}|\Psi_0).
\label{13}
\end{equation}
As anticipated in the Introduction, the form of Eqs. (\ref{10})
is strictly related to the degree of truncation of the boson Hamiltonian.
In the hypothesis that $H_B$ contains up to two-boson terms, the double
commutators in Eqs. (\ref{11}) and (\ref{12}) are just numbers which, therefore,
are also the values of the matrices $A$ and $B$. This is the simplest
case which can be realized in this formalism and represents the boson
counterpart of the standard RPA. This degree of approximation can be improved
by introducing a Hamiltonian with higher-order terms like, for instance, 
(\ref{7}).
In this case the double commutators are operators. In order to calculate
their expectation values in $|\Psi_0)$ , as required in Eqs. (\ref{11}) and
(\ref{12}), one can express 
the $b$ and $b^{\dagger}$ operators in terms of $Q$ and $Q^{\dagger}$ by
reversing Eq. (\ref{8}) (and its conjugate)
and use the fact that the ground state $|\Psi_0)$ is defined as the vacuum
of the Q's. This procedure gives, however, matrices $A$ and $B$ which 
depend on the X and Y amplitudes and, consequently, equations of motion 
(\ref{10}) which are non linear.
In what follows this non linear extension of RPA will be called ERPA. 

Having determined the X and Y amplitudes within RPA or ERPA, one can express
the Hamiltonian $H_B$ in terms of the operators Q and $Q^{\dagger}$. 
In the case of RPA, namely when the boson Hamiltonian (\ref{7}) is truncated
at two-boson terms only, $H_B$ can be rewritten simply as
\begin{equation}
H_B=E_0+\sum_{\nu}{\omega}^{(\nu )}Q^{\dagger}_{\nu}Q_{\nu},
\label{14}
\end{equation}
where ${\omega}^{(\nu )}$ are the energies solutions of the RPA equations 
(\ref{10}). This Hamiltonian  obviously doesn't mix states with different
phonon numbers and so its eigenstates are pure zero-, one-, ..., m-phonon 
states. For a higher-level truncation in the boson Hamiltonian, like for
instance that of Eq. (\ref{7}), $H_B$ acquires instead the more general form
\begin{eqnarray}
H_B=&&E_0+H_{10}(Q^{\dagger}+h.c.)+H_{11}Q^{\dagger}Q+%
      H_{20}(Q^{\dagger}Q^{\dagger}+h.c.)+\nonumber\\
    &&H_{21}(Q^{\dagger}Q^{\dagger}Q+h.c.)+%
      H_{30}(Q^{\dagger}Q^{\dagger}Q^{\dagger}+h.c.)+\nonumber\\
    &&H_{22}Q^{\dagger}Q^{\dagger}QQ+%
      H_{31}(Q^{\dagger}Q^{\dagger}Q^{\dagger}Q+h.c.)+%
      H_{40}(Q^{\dagger}Q^{\dagger}Q^{\dagger}Q^{\dagger}+h.c.),
\label{15}
\end{eqnarray}
(for simplicity, we have dropped all the indices) where the $H_{ij}$
coefficients are functions of $X$ and $Y$. Also in this case, as in RPA, 
the term $H_{20}$ as well as the non-diagonal terms $H_{11}$ vanish, as can be 
easily shown using the fact that the $X$ and $Y$ amplitudes are solutions
of the ERPA equations (\ref{10}). 
The remaining terms of (\ref{15}), however, mix states
with different phonon numbers so that the eigenstates of the full Hamiltonian
become  combinations of these states. This fact introduces an evident 
difference  with RPA since the energies which result from the ERPA equations
are not eigenvalues of the boson Hamiltonian in the phonon space as it is 
in the case of RPA (where they provide the excitation energies of the one-phonon
eigenstates).
In the next section, we will show the results obtained by diagonalizing
(\ref{15}) in different bases, containing up to two-, three- and four-phonon
states, and compare them with the RPA and the exact ones. 

\section{Results and discussion}

Let us first of all fix the parameters entering in the Hamiltonian (\ref{3}).
The energies $\epsilon (1)$ and $\epsilon (2)$ have been chosen equal to
$\epsilon$ and 2.5$\epsilon$, respectively. 
In order to simplify the analysis, the coefficients $V_x(i,j)$, $V_v(i,j)$,
$V_y(i,j,k)$ have been assumed independent of the indices $i, j, k$ and
proportional to one parameter, $\chi$. More precisely, we have fixed
$V_x=-\chi$, $V_v=\frac{1}{4}\chi$, $V_y=-\frac{3}{8}\chi$ as in 
Ref. \cite{gra}.
Both $\epsilon$ and $\chi$ are parameter expressed in units of energy.
All calculations have been performed for a system of $2\Omega =10$ particles.

We begin our analysis by examining the limits of validity of the truncation 
in the boson Hamiltonian (\ref{7}). To do that we compare the exact spectrum 
of the fermion Hamiltonian (\ref{3}) with that obtained by diagonalizing 
$H_B$ (\ref{7}) in the full boson space.
The results are shown in Fig. 1 where 
we report, as functions of the strength $\tau =\frac{2\Omega\chi}{\epsilon}$, 
the energies of those states which,
at zero strength, are pure 1p-1h or 2p-2h states. For later use we also 
label these states with the symbols $|\nu )$ and $|\nu _1\nu _2 )$ meaning
by that the states which, in the limit $\tau\rightarrow 0$, are the
one-phonon $(|\nu )\equiv Q^{\dagger}_{\nu}|\Psi_0))$ and two-phonon
$(|\nu _1\nu _2)%
\equiv Q^{\dagger}_{\nu _1}Q^{\dagger}_{\nu _2}|\Psi_0))$ states of the
RPA formalism. The eigenvalues of the boson 
hamiltonian (dashed lines) are found in good agreement with the exact 
ones in the whole range of variation of
$\tau$. In the following we will take the spectrum of the
truncated hamiltonian as the reference one. 

In Fig. 2, we show the energies resulting from the RPA and ERPA equations
together with the reference spectrum. 
For small values of the strength, RPA and ERPA give almost identical results,
in good agreement with the reference ones. Differences become really
visible around $\tau =0.28$, where RPA undergoes a collapse. In this region
ERPA solutions remain instead stable and close to the reference ones.
On the basis of the comments at the end of Sec. II, however, we notice
that these ERPA energies can be considered excitation energies of
one-phonon states only in the hypothesis that all the terms of (\ref{15})
but $E_0$ and $H_{11}$ play a minor role.

Although not clearly visible at a first glance, the spectrum of Fig. 1
shows marked anharmonicities. In order to quantify these, in Fig. 3
we plot the ratios
\begin{equation}
R_{\nu _1\nu _2}=\frac{E_{\nu _1\nu _2}-(E_{\nu _1}+%
E_{\nu _2})}{E_{\nu _1}+E_{\nu _2}},
\label{16}
\end{equation}
where by $E_{\nu}$ and $E_{\nu _1\nu _2}$ we mean the exact energies
of the states $|\nu )$ and $|\nu _1\nu _2)$ defined in Fig. 1. One notices
a well different behaviour of the three ratios plotted. 
In the case of the state $|11)$ the ratio is seen to grow rapidly for
increasing $\tau$. In the case of the state $|22)$, 
$R_{\nu _1\nu _2}$ remains instead almost
exactly zero therefore showing that the corresponding exact state can be fairly
well described as a pure two-phonon state. The remaining ``mixed'' state
$|12)$ exhibits only a moderate anharmonicity.
These results resembles that of Ref. \cite{two} where a realistic two-electron 
system was examined. Indeed, in both cases, the spectrum is found to
exhibit some levels with a clear harmonic nature mixed, however, to other
levels which do not display at all this nature. This mix of states makes
of course complicate to provide a clear cut answer to the question of how much
harmonic is the spectrum of these many-body systems, the answer strongly 
depending on which levels one is looking at.

The existence of the anharmonicites evidenced in Fig. 3 represents an
evident limit to the harmonic picture of RPA and points to the need of
diagonalizing the full Hamiltonian (\ref{15}) in a multiphonon space
in order to reproduce the exact spectrum properly.
We have performed three separate calculations
by diagonalizing $H_B$ in spaces including up to two-, three- and four-phonon
configurations. The results are shown in Fig. 4 together with the reference
ones. 
The internal structure of the collective phonons entering in the basis states
is the one obtained in ERPA, i.e. the $X$ and $Y$ amplitudes are solutions
of Eqs. (\ref{10}) where the full Hamiltonian (\ref{7}) has been used.
Keeping for simplicity the same labelling of states used in Fig. 1, 
we can say that,
as expected from the previous discussion, for the states $|1)$, $|2)$ and
$|22)$ the results are
almost independent of the basis and in very good agreement with the exact 
ones. For the state $|11)$ we find instead a marked dependence
on the basis, the results approaching more and more the reference ones when
the space is enlarged. Going from the smaller space 
to that including up to three
phonons the agreement is quite good for $\tau\alt 0.25$. For a higher strength
the space needs to be further enlarged up to four phonons. 
A somewhat similar behaviour, although less pronounced, is seen for the
state $|12)$.

As a final result, in Fig. 5 we show the ground state energies calculated
within RPA and as they result by diagonalizing the Hamiltonian
(\ref{15}) in multiphonon spaces. All these results are
compared with the exact ones and with those obtained by diagonalizing
the boson Hamiltonian (\ref{7}). It is interesting
to see that the good agreement between fermion and boson results
already found in Fig. 1 for the excitation energies is confirmed also 
in this case. As expected, RPA overestimates the exact values. 
All the diagonalizations in multiphonon spaces 
lead instead to an underestimation of these values showing, however, 
a rapid convergence to the exact values with increasing the size of
the spaces.

\section{Conclusions}
In this paper we have analysed some of the lowest excited states of the
spectrum of a solvable 3-level model, namely those that in RPA would
be described as one and two-phonon states. We have worked in a boson 
formalism. As a preliminary step, then, we have constructed a boson
image of the fermion Hamiltonian whose expansion has been truncated at 
four-boson terms. The procedure followed in such a derivation has been 
of Marumori-type. The quality of the boson Hamiltonian has been tested
by comparing its eigenvalues with the exact ones. Within the considered
range of
variation of the interaction strength the agreement between fermion and
boson energies has always been found quite good. 

By making use of this boson Hamiltonian we have constructed RPA-type 
equations and solved them iteratively. The new solutions have gained in
stability with respect to the RPA ones and, in particular, around the RPA
collapse point the new energies have exhibited a good agreement with the
exact ones. This extension of RPA, while introducing corrections to
the Pauli Principle violations present in RPA, naturally leads to a
Hamiltonian which mixes states with a different number of phonons. We
have performed diagonalizations in spaces containing up to two-, three- and
four-phonon states and observed an improved quality of the 
approximate spectra with increasing the size of the spaces.

Special attention has also been addressed to the problem of the anharmonicity 
of the spectrum. This has been found relevant for the state which, in RPA, is
described as the double excitation of the lowest one-phonon state. On
the contrary, for the other states, a less pronounced anharmonicity has
been found. These findings agree with those of Ref. \cite{two} and
point to the necessity of considering together all the possible
elementary excitations of a many-body system when discussing the 
anharmonicity of its spectrum.

\newpage

\begin{figure}
\caption{Excitation energies of the states $|\nu>$ and $|\nu_1 \nu_2>$
which, at zero interaction strength, 
are pure 1p-1h and 2p-2h states respectively. The
exact energies (full lines) are compared with the energies obtained
diagonalizing the hamiltonian $H_B$ in the full boson space (dashed lines).
The energies are shown as functions of the parameter $\tau$.}
\label{1a}
\end{figure}

\begin{figure}
\caption{Excitation energies, as functions of $\tau$, 
calculated within RPA (dotted lines) and ERPA
(dashed lines) compared with the corresponding 
reference energies (full lines).}
\label{2a}
\end{figure}

\begin{figure}
\caption{The ratios $R_{\nu_1 \nu_2}$ (Eq.(\ref{16}))
 calculated for the states $|11>$, $|12>$ and $|22>$ as functions of $\tau$.}
\label{3a}
\end{figure}

\begin{figure}
\caption{Excitation energies, as functions of $\tau$, obtained diagonalizing
$H_B$ in spaces including up to two- (dotted lines), three- (dot-dashed lines)
and four- (dashed lines) phonon configurations compared with the reference
excitation energies (full lines).}
\label{4a}
\end{figure}

\begin{figure}
\caption{Ground states energies, as functions of $\tau$, obtained in the
exact fermionic calculation (full line), in the reference calculation (dashed
line), in RPA (+ symbols), in the diagonalizations up to two phonons (left
triangles), up to three phonons (diamonds) and up to four phonons (circles).}
\label{5a}
\end{figure}

\newpage

\includegraphics{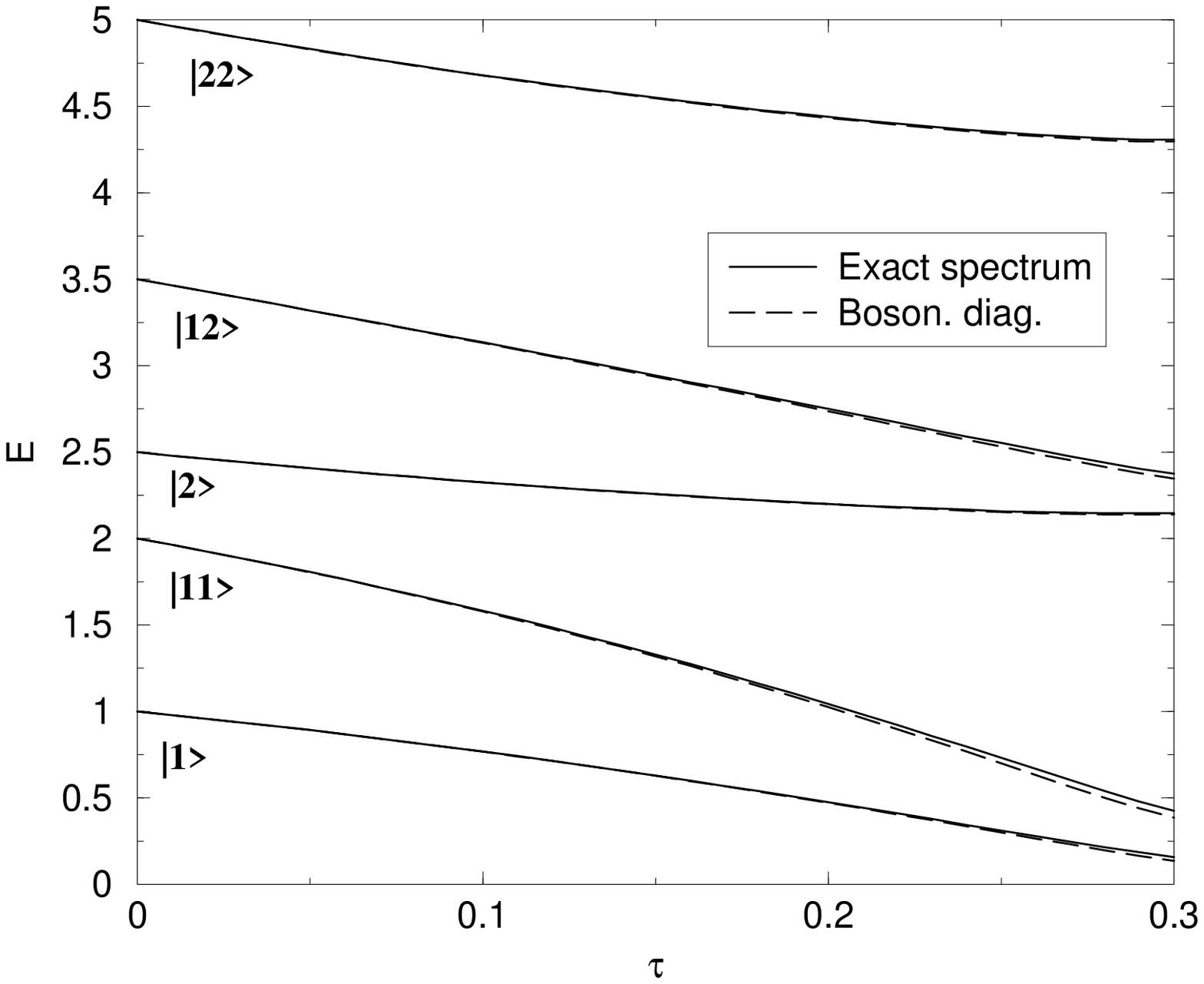}
\mbox{}\\

\newpage

\includegraphics{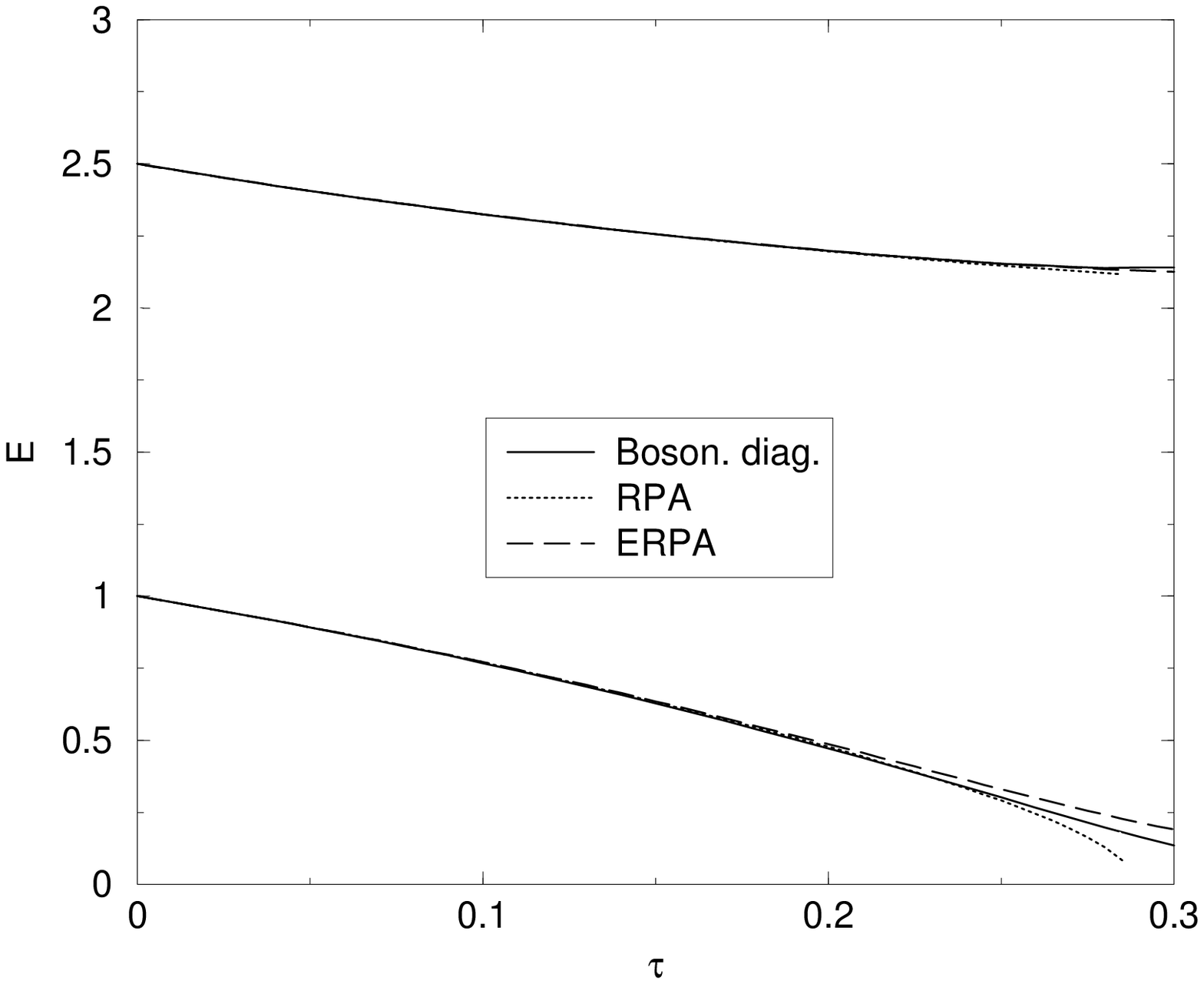}
\mbox{}\\

\newpage

\includegraphics{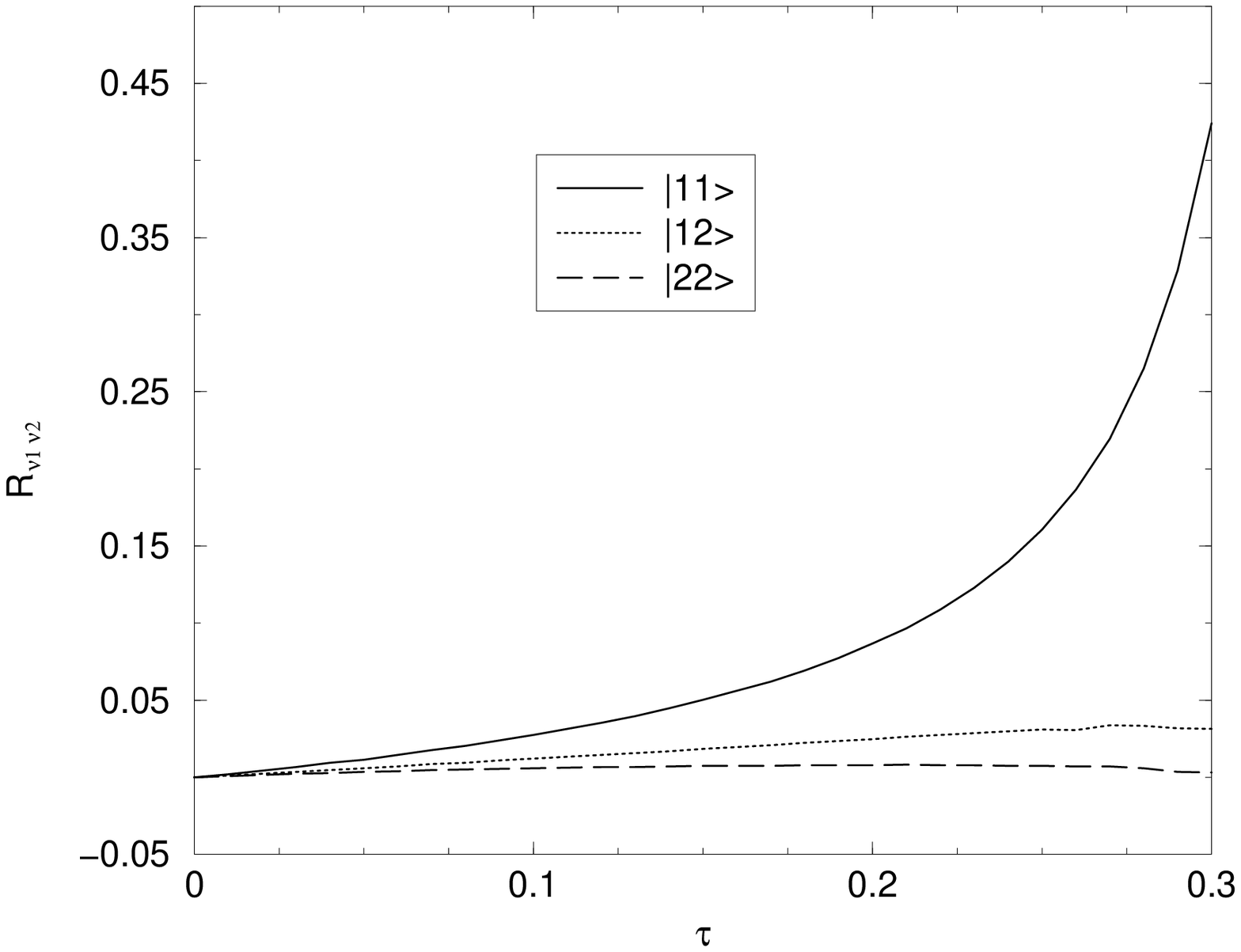}
\mbox{}\\

\newpage

\includegraphics{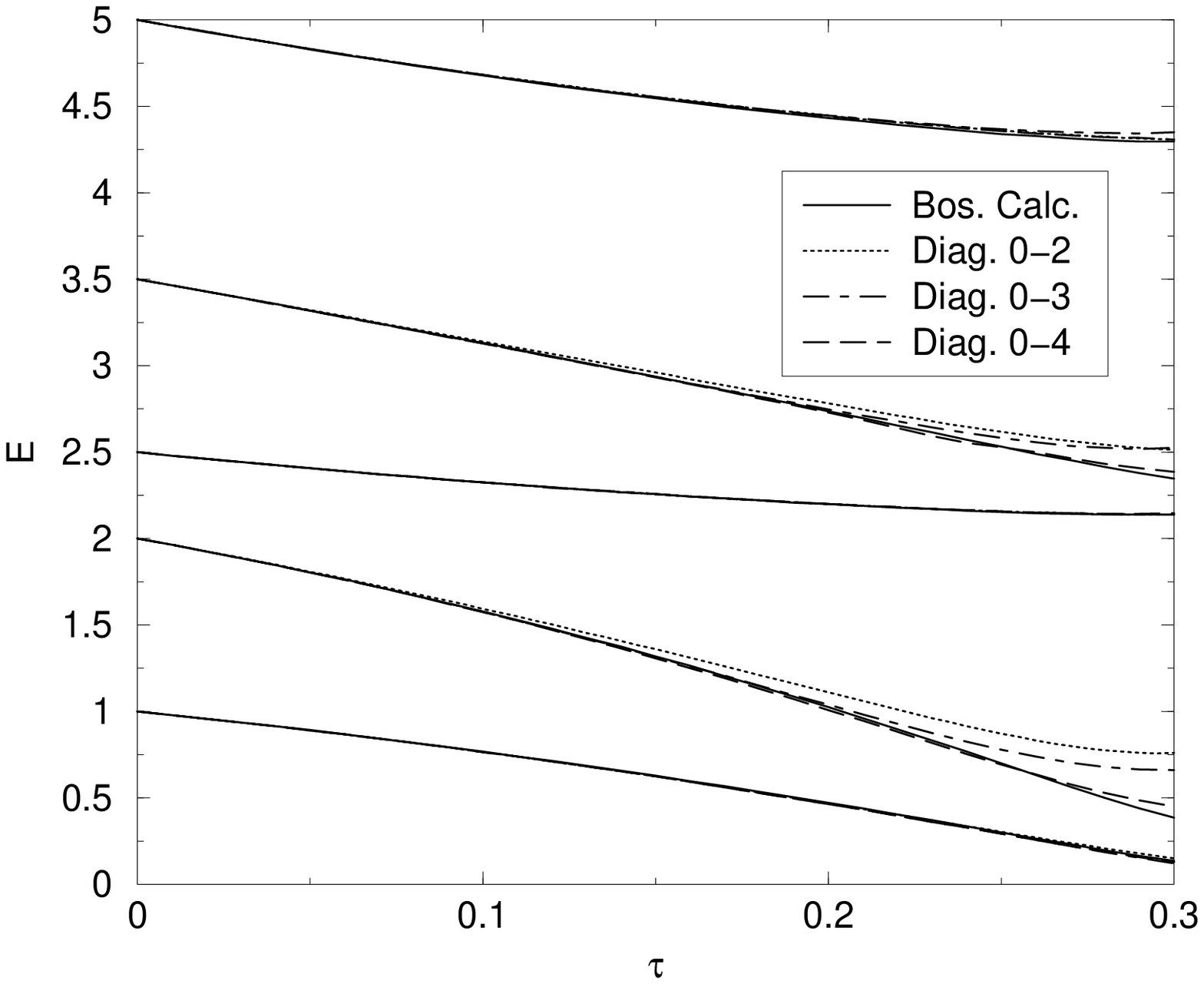}
\mbox{}\\

\newpage

\includegraphics{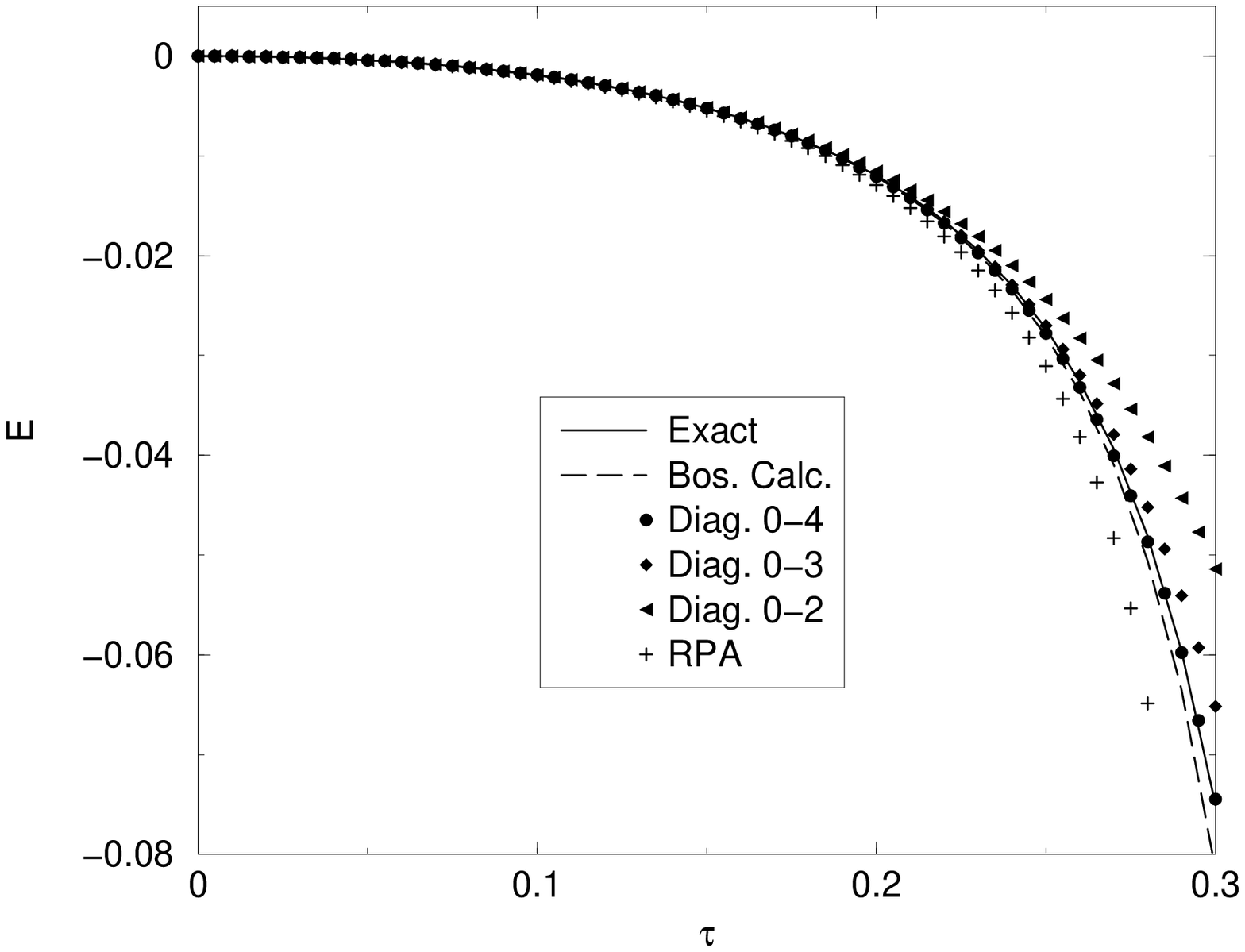}
\mbox{}\\

\end{document}